\documentclass[11pt]{article}
\usepackage{a4wide,parskip,appendix}
\usepackage{amsfonts}
\usepackage{pdfpages}
\usepackage{mdframed}
\usepackage{titlesec}
\usepackage{float}
\usepackage{venndiagram}
\usepackage{hyperref}

\usepackage{amsmath}

\usepackage{hyperref}
\usepackage{caption}
\usepackage{algorithm2e}

\usepackage{xcolor}
\definecolor{LRcolour}{RGB}{200,250,200}
\definecolor{Dcolour}{RGB}{210,230,250}
\definecolor{MixColour}{RGB}{250,230,200}

\usepackage{csquotes}
\MakeOuterQuote{"}

\usepackage[a4paper, total={6in, 9in}]{geometry}

\titlespacing*{\section}
{0pt}{5.5ex plus 1ex minus .2ex}{4.3ex plus .2ex}
\titlespacing*{\subsection}
{0pt}{5.5ex plus 1ex minus .2ex}{2.3ex plus .2ex}

\RestyleAlgo{ruled}

\begin{document}

\title{\textbf{Gaming and Blockchain: Hype and Reality}}
\author{Max McGuinness (mgm52), Queens' College \\ \emph{R47 - Distributed Ledger Technologies}}
\maketitle

\begin{abstract}
This paper explores the adoption of blockchain technology in the gaming industry. While supporters affirm that distributed ledger technology has potential to revolutionize gaming economies and provide players with control over their virtual assets, there are practical challenges such as energy consumption and user adoption to be addressed, and detractors question whether blockchain integration is even necessary. This report characterises popular blockchain-based gaming projects like Enjin and Axie Infinity, then goes on to compare metrics such as transaction cost and player feedback to evaluate the longevity of blockchain-integrated gaming as a whole.

\tableofcontents

\end{abstract}

\newpage
\vspace*{\fill}
\begin{center}
\includegraphics[width=7.5cm]{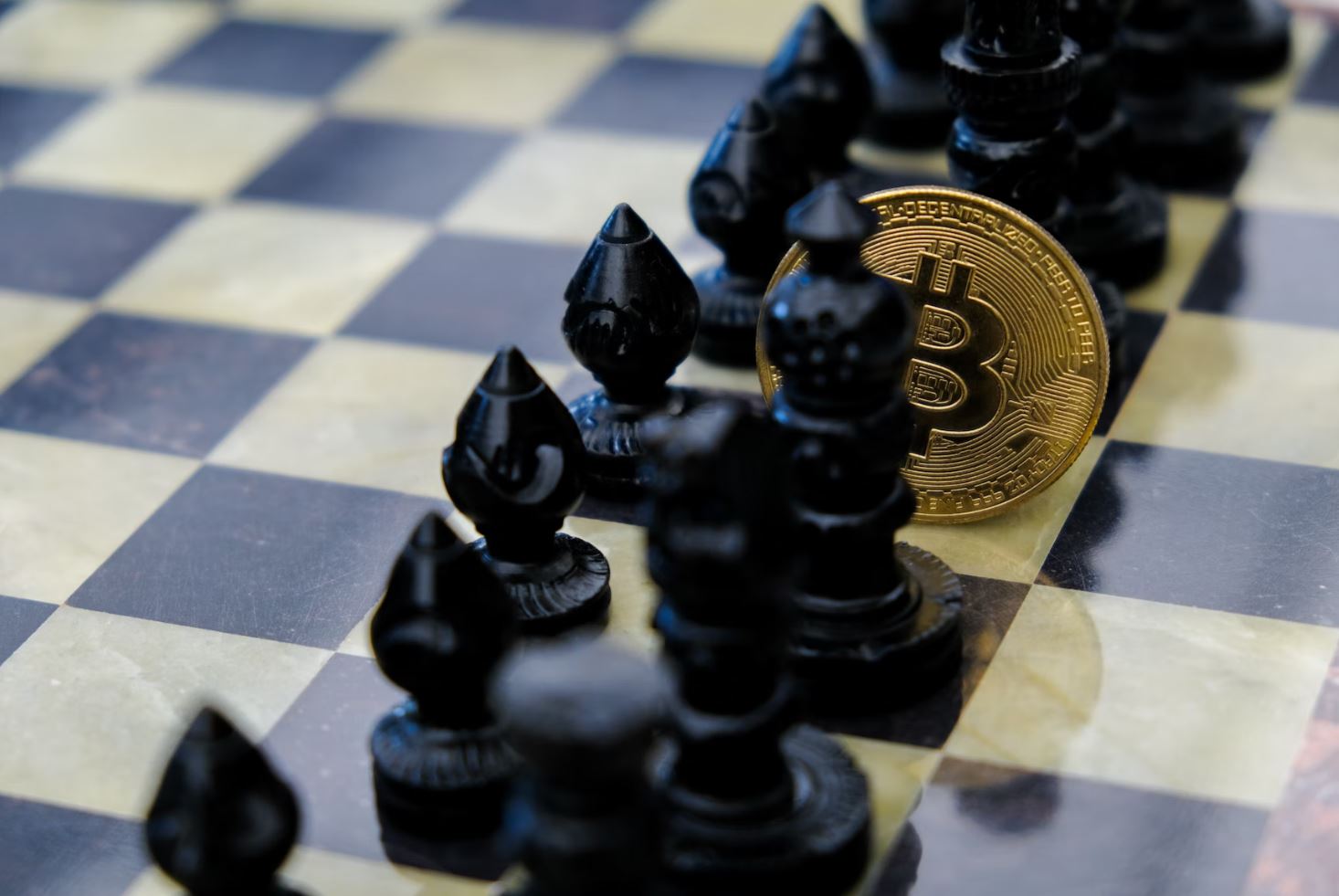}
\captionsetup{width=0.5\linewidth}
\captionof{figure}{\footnotesize An unrealistic depiction of blockchain-gaming integration. \textit{Image used freely under the Unsplash License}.}
\label{fig:chess}
\end{center}
\vspace*{\fill}
\newpage

\section{Introduction}
This paper focused on the integration of blockchain into the gaming industry. That is to say, it focused on projects which are compelling games in their own right, and interface with a blockchain for any reason. I refer to these as \textbf{blockchain gaming} (BG) projects. I aim to present a thorough discussion of the core components of blockchain gaming as it has evolved since the mid-2010s, both by examining key projects and looking at broader trends.

\subsection{Blockchain Technology}
Dating back to the late 1970s, a blockchain is a means of maintaining a distributed timestamped ledger, in which older transactions are effectively validated by the presence of newer ones \cite{first_blockchain}. Nakamoto's key innovation in inventing Bitcoin was to describe a \textit{decentralised} blockchain, which they achieved by repurposing proof-of-work for \textit{consensus} \cite{nakamoto_2008}. While proof-of-work had existed as a concept since the early 90s (primarily for denial-of-service prevention \cite{first_proof_of_work} \cite{hashcash}), this was the first time the algorithm had been used in such a way. Modern usage of the term \textit{blockchain technology}, particularly in the context of gaming, tends to refer to it in decentralised scenarios like the one described by Nakamoto.

Blockchain technology was again revolutionised in 2014, when Vitalik Buterin's Ethereum blockchain reduced barriers to creating decentralised applications (\textit{dApps}) via pro-development features like turing-complete smart contracts and stateful storage \cite{buterin2016ethereum}. Ethereum was the foundational technology for blockchain games, and to this day a significant proportion run on ETH token standards. The most important Ethereum standards for the sake of this paper are:
\begin{itemize}
    \item ERC-20, for simple fungible tokens.
    \item ERC-721, for non-fungible tokens (NFTs).
    \item ERC-1155, for multi-tokens transactions (including mixing fungible \& non-fungible).\footnote{Notably, Enjin is an ecosystem of tools to help blockchain integration in games by utilising ERC-1155 assets.}
\end{itemize}

\subsection{Categorising Approaches}
Several existing studies \cite{vidal-tomás_2022, kanchanaratanakorn_chutima_2022} categorise blockchain gaming projects into two approaches: \textbf{Play-to-Earn} and \textbf{Metaverse}, likely due to the categories existing within the top 20 on \hyperlink{https://www.coingecko.com/en/categories}{CoinGecko's Crypto Categories By Market Cap}.\footnote{As of 2023/03/25.} I personally find this categorisation unhelpful: while it may be informative to those tracking cryptocurrency opportunities in an investor-facing app such as CoinGecko, it seems that the most popular BG projects fall largely into \textit{both} Play-to-Earn and Metaverse descriptions. For example, consider Axie Infinity, a virtual world in which cryptocurrency is awarded to players for winning battles \cite{axie-infinity}; or VoxVerse, in which virtual avatars compete to earn TownCoins \cite{voxverse_2023}.

The term "Metaverse" is particularly poorly defined. At its most basic level, one could extend its definition to apply to all multi-player 3D worlds \cite{metaverse-poorly-defined}, and Wang et al point out that almost any practical NFT application can be described as having potential Metaverse usage \cite{nft-applications}. The dual categorisation also fails to address important differences such as the use of fungible or non-fungible tokens, custom or pre-existing blockchain, and ability to trade assets in-game or on third-party markets.

One should also consider how the line between categorising a project as primarily \textit{finance-} or primarily \textit{gaming-}focused can be blurred. Axie Infinity co-founder Aleksander Larsen has stated that "\textit{GameFi, as a definition, means gamified finance. Blockchain gaming means something totally different.}"\cite{gamefi-definition} As an example, he described a \textbf{GameFi} system in which users can gain experience and "level up" whenever they perform transactions on a particular protocol (a gaming system \textit{enhancing} a DeFi project). By contrast, he uses the term \textbf{Blockchain Gaming} to refer to projects built primarily to be games from the ground-up (a blockchain protocol \textit{enhancing} a compelling game).

While this paper primarily focuses on Larsen's \textit{Blockchain gaming} categorisation, the two concepts are very much intertwined.

\section{Real-World Approaches}

\subsection{Token-Oriented Integration}
Overwhelmingly, blockchain-game integration comes in the form of distributing tokens to players: either representing in-game assets (Section \ref{sec:ingame-markets}), or simply as an extrinsic reward (seen with \textit{Play-to-Earn} and its derivatives, Section \ref{sec:play-to-earn}). Hypothetically, one could use blockchain technology for mechanical purposes other than managing tokens, which I mention in Section \ref{sec:nontoken}.

\subsubsection{Play-to-Earn} \label{sec:play-to-earn}
The play-to-earn (P2E) model can be broadly categorised as awarding players with blockchain assets (fungible or non-fungible) for engagement with a game. Compilation websites such as \texttt{playtoearn.net} focus on the idea of it as a source of free income; detractors says it is a proxy for ponzi schemes due to constant inflation.

A particularly pure example would be \textbf{RollerCoin}, which is notable for both its initial lack of NFTs and the fact that players primarily earn cryptocurrencies not directly controlled by the developers \cite{rollercoin}.\footnote{Though this point is made slightly moot by the fact that the cash-out rate is still controlled by the developers. This means that, for players, the process of earning Bitcoin on their platform is fairly equivalent to acquiring some quantity of ERC-20 token, as you would in another play-to-earn game, and then trading it for Bitcoin on a third-party market.} In RollerCoin, players complete minigames that simulate the process of Bitcoin mining (but don't run real mining operations), and eventually cash out in their choice of Bitcoin, Ethereum, or Dogecoin. It is possible to play and earn entirely for free, though non-paying non-referring players experience low revenue, being at a maximum of around \$0.30/day according to some estimates \cite{rollercoin_earnings}. The developers claim that the project is funded by sponsorships, in-game advertising, and the purchase of in-game items, enabling the funding of free players \cite{rollercoin}. The top-earning player sits at around \$50/day.

An important subcategory of play-to-earn is \textbf{move-to-earn}, which awards players for physical exercise through use of GPS tracking, such as via a smart watch or phone. Cryptorank called it the "biggest GameFi trend of 2022" \cite{rise-and-fall-of-play-to-earn-games}, particularly highlighting the STEPN app that awards players with \textit{Green Satoshi Tokens} and was the progenitor to many move-to-earn games \cite{stepn}. Interestingly, like RollerCoin, STEPN's move-to-earn mechanics exist alongside an NFT marketplace; though unlike RollerCoin, STEPN utilises the Solana blockchain. STEPN claim they chose Solana because of how "\textit{cheap, smooth, and seamless it is}" \cite{stepn_uses_solana}.

In fact, many \textit{play-to-earn} games also make use of an in-game market (Section \ref{sec:ingame-markets}). \textbf{Axie Infinity} \cite{axie-infinity} was one of the most popular examples, in which players can breed, battle and trade fantasy creatures called Axies. Developer Sky Mavis charge a 4.25\% fee to players when they trade Axies on its marketplace. The play-to-earn branding related to the idea that playes could earn the game's ERC-20 tokens \textit{AXS} and \textit{SLP} by simply playing the game, after paying some initial starting cost.

\begin{figure}[h]
\begin{center}
\includegraphics[width=0.48\linewidth]{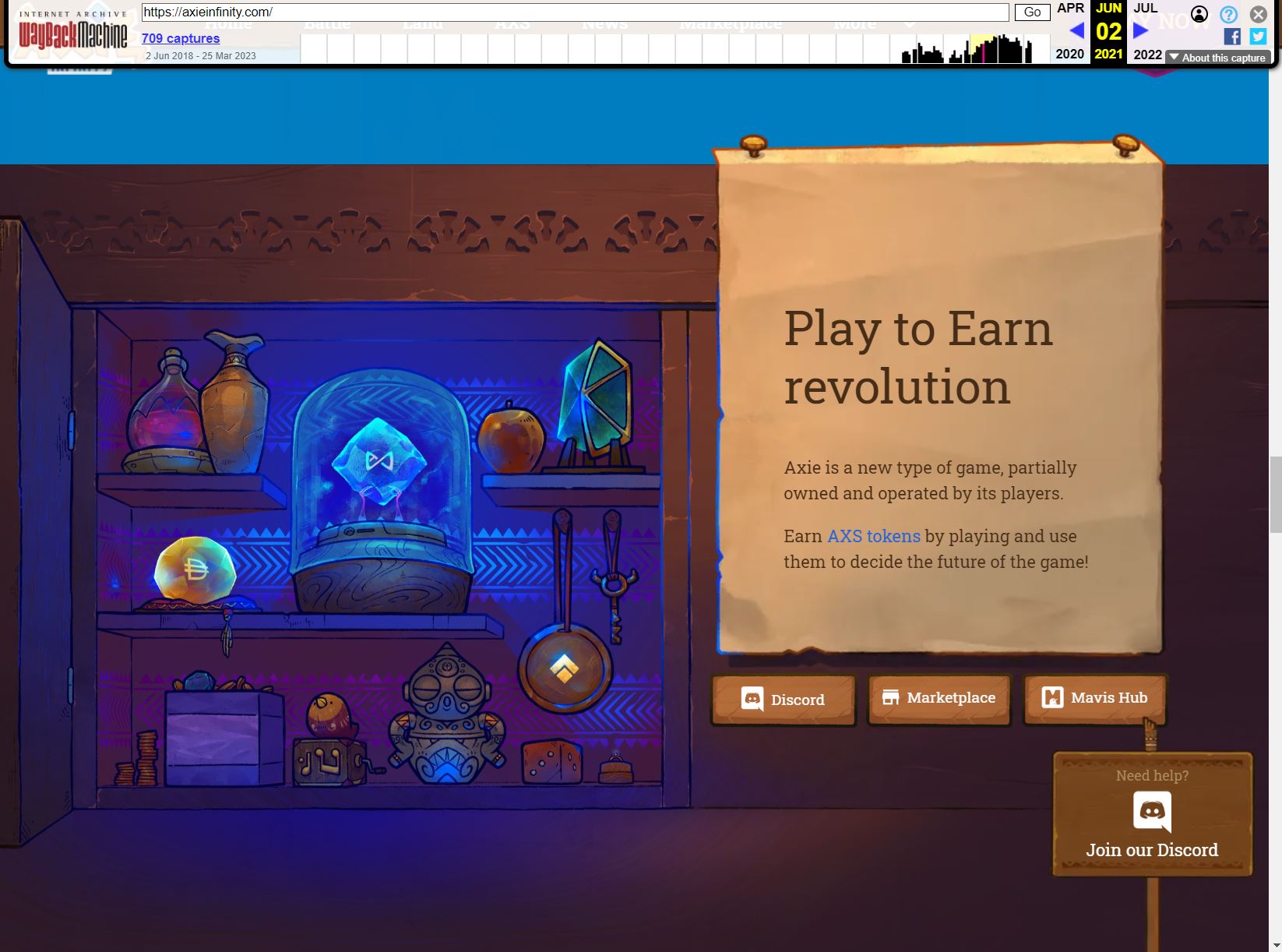}
\hfill
\includegraphics[width=0.48\linewidth]{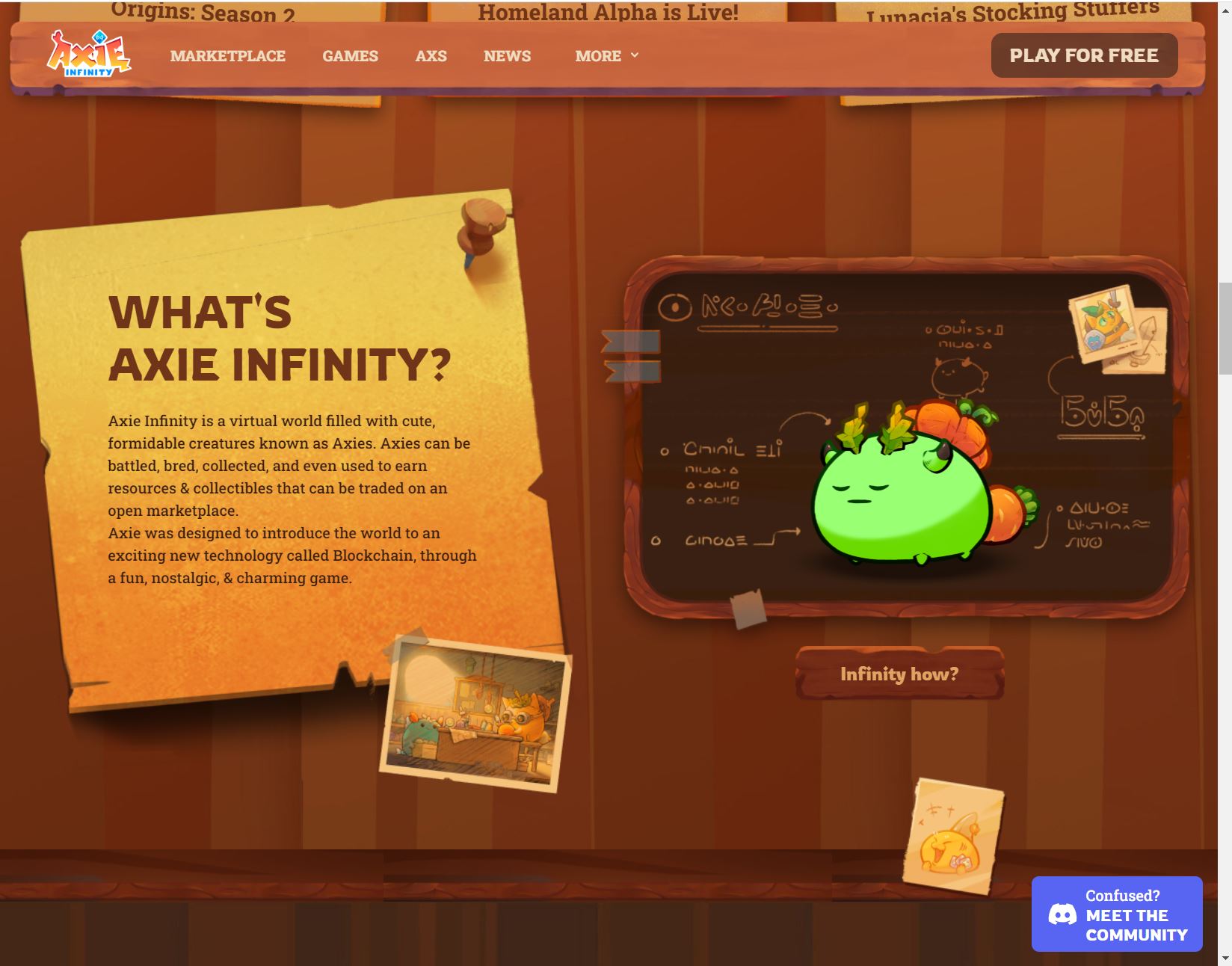}
\end{center}
    \caption{\small Left: an archival screenshot of the Axie Infinity website on June 2021. Right: the Axie Infinity website today (2023/03/25), showing the removal of its \textit{play-to-earn} branding.}
    \label{fig:axie_p2e}
\end{figure}

Despite its pay-to-play (Section \ref{sec:p2p}) starting cost being as high as \$400 in February 2020, some users in the Philippines were reportedly able to treat the game as their main source of income by June 2021 owing to the climbing value of SLP \cite{people_lived_on_p2e}. While the matter is subjective, I believe that it would not be amiss to compare the promise of future returns on an initial investment to the operation of a \textit{Ponzi scheme}.\footnote{In fact, there may be an argument to be made that it is inherently impossible for a game to be both \textit{pay-to-play} and \textit{play-to-earn} without having characteristics of a Ponzi scheme, due to the hypothetical syllogism of \textit{\textbf{pay-to-earn}}.} In February 2022, as the game declined in popularity and the wider NFT market experienced a crash, the value of SLP plummeted to less than 1\% of its peak value. This caused a mass exodus of players, and Sky Mavis tactfully removed the phrase "play-to-earn" from its branding on all media \cite{axie_infinity_was_a_bad_p2e_bubble}. This is demonstrated in Figure \ref{fig:axie_p2e}.

\subsubsection{Pay-to-Play} \label{sec:p2p}
If the \textit{play-to-earn} model can be characterised as distribution of tokens from developers to players, \textit{cryptocurrency payments} act in the opposite direction: permitting players to pay the developers for items using cryptocurrency. I term this "\textit{pay-to-play}". In reality, many games that have one model will also have the other - but they are nonetheless worth separate examination.

One of the first gaming companies to accept Bitcoin as a legitimate payment option was \textbf{BigFish Games}, who added Bitcoin support for purchase of all of their games, and some in-game transactions, in early 2014 \cite{bigfish_btc_games}. \textbf{Microsoft} also became an early adopter in 2014 when it began accepting BTC as payment to buy games (and other other digital content) in the Microsoft Store, using the \textbf{Bitpay} \cite{bitpay_2014} platform as a payment processor \cite{microsoft_accepts_btc}. A Microsoft representative described their motivation as being "about giving people options and helping them do more on their devices and in the cloud" \cite{microsoft_accepts_btc}.

The acceptance of cryptocurrency is not without controversy, however, and there has been some debate over the use of it as a payment method due to its volatility and lack of regulation \cite{hard_to_use_crypto_payments}. \textbf{Steam} dropped support for cryptocurrency payments in December 2017 such reasons, stating that Bitcoin's volatility and high fees made support untenable, as there were price fluctuations even while customers were at check-out \cite{steam_doesnt_accept_bitcoin}.

\subsubsection{In-Game Markets} \label{sec:ingame-markets}
While \textit{play-to-earn} and \textit{pay-to-play} paradigms sometimes see their blockchain implementation exist purely outside the confines of the game world (i.e. as a player-developer interaction, not a player-game interaction), in-game markets embed the idea of tokens directly into the gameplay itself, making the exchange of tokens part of the intended experience.

The ERC-721-based \textbf{CryptoKitties} \cite{cryptokitties_2023} is the archetypal example. Launched in late 2017, CryptoKitties is emblematic of the "breeding" subgenre of games, in which player-owned assets can be combined (at a cost) to create new ones; this is a popular mechanic amongst BGs for giving NFTs an easily-defined utility and discouraging them from being sold \cite{rise_and_fall_cryptokitties}. Aside from breeding, players can purchase or trade randomly-generated NFT "kitties" at the cost of Ethereum.

CryptoKitties was the precursor to other tropes of token-oriented BGs, too, such as card-collecting, randomised stats, rarity tiers, and the use of a web interface - at a time when web games were declining in popularity amid the death of Flash \cite{browser_game_decline}. These mechanics have all existed long before BGs - for example, with breeding being traced back to Pokemon, and item tiers to Diablo - but CryptoKitties' exploding popularity established them as core tools for modern BG developers.\footnote{Examples of derivative BGs would include \textbf{Etheremon} \cite{ethermon} (with more involved gameplay than CryptoKitties), \textbf{EtherGoo} \cite{ethergoo} (buy cats to produce resources), and even \textbf{KotoWars} \cite{kotowars} (battling game which uses players' existing CryptoKitties NFTs).}

The seemingly influential nature of CryptoKitties as a progenitor of BGs may be an example of its \textit{first-mover advantage}. Being able to brand itself as "the world’s first Ethereum game", CryptoKitties experienced viral popularity after launch, spiking from 1.5k NFT sales on launch day to over 50k/day in mid-December of 2017. The movement was so large that it even provoked a spike in Ethereum's value, accounting for 25\% of ETH traffic at the game's peak and influencing ETH's movement from \$300 to \$1,400 between late 2017 and early 2018. Of course, this popularity was ephemeral and has since been categorised as one of the many examples of bubbles in the crypto space \cite{cryptokitties_was_a_bubble}. CryptoKitties went on to experience fewer than 100 NFT transactions per day on average in September 2022 \cite{cryptokitties_2023}.

\subsection{Non-Token-Oriented Integration} \label{sec:nontoken}
While some token-oriented games will also take advantage of the blockchain's transparency beyond the tokens themselves, such as to prove breeding rules in smart contracts \cite{cryptokitties_2023}, games that use blockchain technology primarily for mechanisms \textit{other} than token-representation are relatively few and far between.

Perhaps the most prominent example of a game mechanic that can benefit from a blockchain implementation is \textit{randomisation}. This was even referenced in the original Ethereum whitepaper \cite{buterin2016ethereum}, which cited Frank Stajano and Richard Clayton's \textbf{Cyberdice} proposal for transparently computing gambling outcomes in a peer-to-peer setting \cite{cyberdice}. Today, blockchain-implemented gambling games are widespread, with popular examples including \textbf{BC.Game} \cite{bc.game}; \textbf{MyStake} \cite{stake}; and \textbf{Bets.io} \cite{bets.io}; though these may be stretching my definition of \textit{non-token-oriented integration}.

The legal status of cryptocurrency-based gambling/casinos varies across the world, in many cases remaining a grey area. In some countries, foreign online casinos are widely accessible and local regulators do not apply any legislative requirements. However, in others, online gambling is only allowed with a local license, and different restrictions might apply to blockchain or cryptocurrency-based sites. For example, in the UK, the Gambling Commission has been continually updating their expectations on key event reporting for the use of crypto-assets such as BTC or ETH \cite{uk_gambling_commission_btc}. Notably, in the US, cryptocurrency-based casinos have been able to sidestep restrictions in some states where online gambling is banned, owing to the difficulty of regulating the currencies' movement \cite{btc_gambling_loophole}.

\subsection{Integration into Existing Games} \label{sec:existing}

In the years since Cryptokitties exploded into popularity in 2017-18 \cite{cryptokitties_2023}, a number of prominent developers have expressed interest in integrating blockchain technology into existing games; typically in the form of limited-edition collectable items.

For example, in December 2021, Ubisoft announced its first foray into the world of cryptocurrency with the launch of \textbf{Ubisoft Quartz} \cite{ubisoft_quartz}, a program allowing users to purchase NFTs representing in-game cosmetic items, nicknamed "Digits". Based on the Tezos proof-of-stake blockchain \cite{goodman2014tezos}, a strong marketing emphasis was on their environmentally friendly nature, quoting that one Quartz transaction consumes "\textit{1 million times less energy than a Bitcoin transaction}"\cite{ubisoft_quartz}.

Interestingly, this platform aimed to exist \textit{on top of} existing (centralised) in-game cosmetic stores in Ubisoft games, rather than replacing them. Initially launched exclusively for the open-world multiplayer shooter \textbf{Ghost Recon Breakpoint}, the intent was that players would acquire Digits on a separate app, and that these could be re-sold on third-party marketplaces for cryptocurrency. A key selling point of the items was their uniqueness: each would be tagged with a unique serial number visible in-game, as well as displaying the names of previous owners according to its transaction history.

Ubisoft's endeavour was received poorly by players. The general sentiment was that the Quartz program was less trust-worthy than a centralised store (perhaps due to many prominent hacking incidents in the crypto space \cite{crypto_hacks}, and proof-of-stake models being notoriously challenging to secure \cite{pos-insecure}); that the benefits of decentralization were defeated by all Digit transactions requiring an eligible Ubisoft account; that there was too little accountability on Ubisoft's part if users were, say, scammed out of their Digits; and lastly that Ubisoft were only integrating a blockchain to capitalise on it as a buzzword, rather than for a practical purpose \cite{ubi_quartz_bad}.

In April 2022, nearly 4 months after Quartz's launch, Ubisoft announced that they would not distribute any further Digits. An analysis by Ars Technica found that, amongst the thousands of Digits released by Ubisoft between December and April, only 96 successful secondhand trades took place \cite{ubi_quartz_failed}. As of 2023/03/25, Ubisoft has not announced any plans for Quartz support beyond Ghost Recon Breakpoint.

\section{Assessment}

\subsection{Developer Opinions}
Developer reactions to gaming-blockchain integration are mixed. Valve and Mojang have committed to banning NFT-based games from their gaming platforms Steam and Minecraft respectively \cite{valve_bans_nfts, mojang_bans_nfts}. An important reason cited by Valve was the idea that blockchain integration could allow developers to side-step gray market rules, as players could trade in-game items for cryptocurrency, and that cryptocurrency for fiat, in a manner that's hard to trace \cite{blockchain_gaming_loophole, btc_gambling_loophole}.

On the other hand, corporations including Epic, Ubisoft and EA may be embracing blockchain integration. EA CEO Andrew Wilson called NFT and “play-to-earn” games “the future of our industry” \cite{ea_likes_nfts}. Ubisoft’s VP of the Strategic Innovations Lab Nicolas Pouard has outlined why he believes players should care about Ubisoft's move into blockchain, stating that "blockchain is a game changer" \cite{ubi_likes_nfts}. Furthermore, in 2021, Sega announced a collaboration with game developer Double Jump.Tokyo that sees the company beginning to sell NFTs based on its gaming IPs \cite{sega_likes_nfts}. Sega said it will use the new initiative to launch various services which will allow NFT owners to “effectively utilise and further enjoy their NFT content” \cite{sega_likes_nfts}.

\begin{figure}[h]
\begin{center}
\includegraphics[width=1\linewidth]{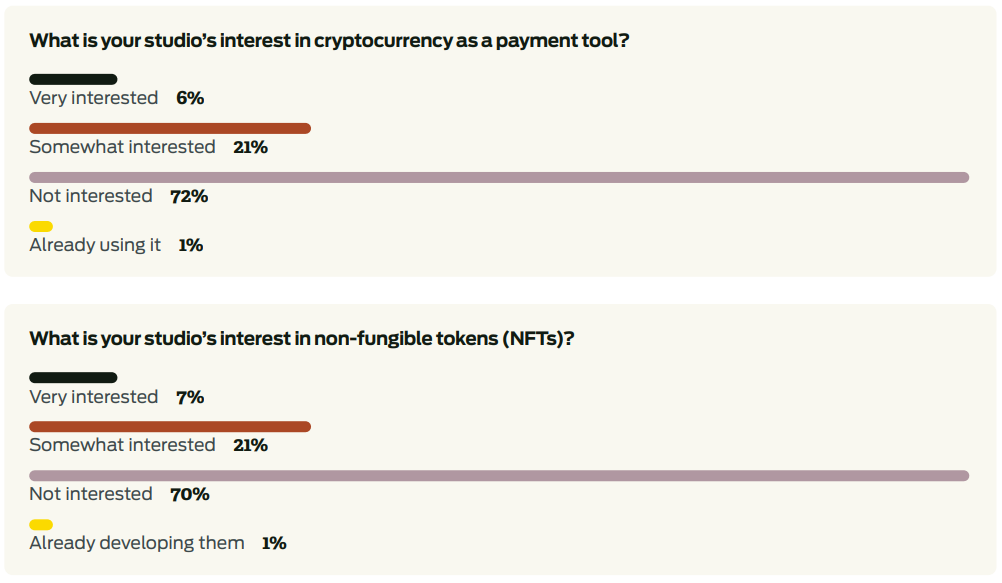}
\end{center}
    \caption{\small A sample of results from GDC's 2022 survey of attendees. According to GDC, when asked how they felt about the idea of cryptocurrency or NFTs in games, a "vast majority" of respondents spoke out against both practices.}
    \label{fig:gdc}
\end{figure}

However, turning our attention to individual developers, we see a more hesitant (or perhaps apathetic) view of adoption in the industry. A 2022 survey by GDC found that as high as 70\% of developers surveyed said their studios had "no interest" in NFTs, while 28\% reported they were "very or somewhat interested" in them, and only 1\% reported their studio to be planning integration into their games \cite{gdc_dev_survey}. I highlight these results in Figure \ref{fig:gdc}.

\subsection{Player Opinions}
\begin{figure}[H]
\begin{center}
\includegraphics[width=1\linewidth]{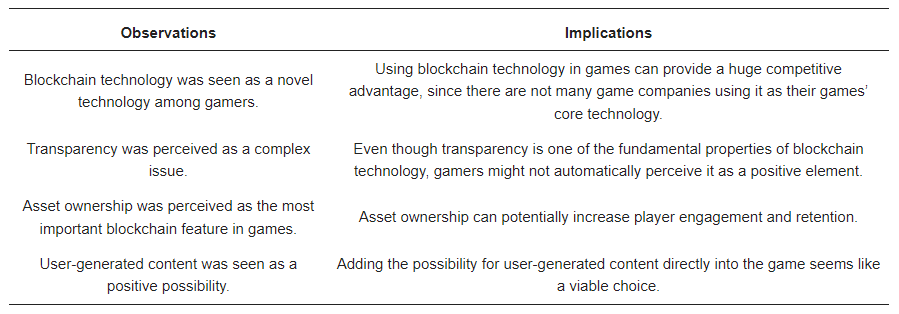}
\end{center}
    \caption{\small Table 6 from Paajala et al's 2022 study, \textit{Users’ Perceptions of Key Blockchain Features in Games} \cite{blockchain_perceptions}.}
    \label{fig:perceptions}
\end{figure}

In an October 2022 study, Paajala et al created a blockchain game and demonstrated it to a number of interviewees, collecting qualitative results through structured interviews. The study found that transparency is one of the core benefits of blockchain-based systems, though many interviewees raised both privacy concerns and possible applications of transparent data. The study also found that blockchain in gaming was not seen as a security threat, though participants were reluctant to comment on whether they trusted blockchain applications ("\textit{not a single participant gave a direct answer to that question}"). Their key findings, summarised through coding, are shown in Figure \ref{fig:perceptions}.

Conversely, however, player reactions to blockchain gaming in-the-wild have often been negative, particularly in the context of integration with existing games (Section \ref{sec:existing}). As a prominant example in AAA gaming, the announcement trailer for Ubisoft Quartz \cite{ubisoft_quartz} reached 96\% dislikes within 24 hours of its release on 7 Dec 2021 \cite{ubisoft_quartz_trailer_hated}. It is currently unlisted on Youtube \cite{ubisoft_quartz_trailer}; the most-viewed public copy of the video is IGN's release, at 124K views and 92\% dislikes as of 2023/03/25 \cite{ubisoft_quartz_trailer_ign}.

The extremity of this reaction does lead me to speculate on whether there might be a kind of \textit{halo effect} on one or both sides of the blockchain-gaming argument, given the enormous investment given by its supporters \cite{cryptokitties_was_a_bubble} and enormous antagonism by its detractors \cite{ubisoft_quartz_trailer_ign}.

\begin{figure}[H]
\begin{center}
\includegraphics[width=0.66\linewidth]{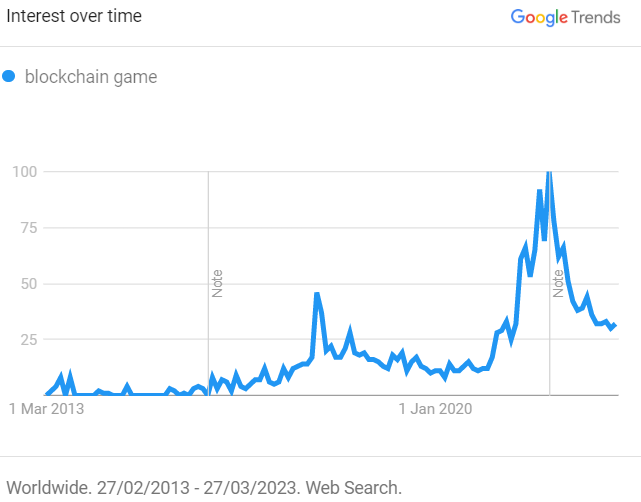}
\end{center}
    \caption{\small Google search queries for "blockchain game" \cite{google_trends_bg}.}
    \label{fig:bg_google_trends}
\hfill
\begin{center}
\includegraphics[width=0.66\linewidth]{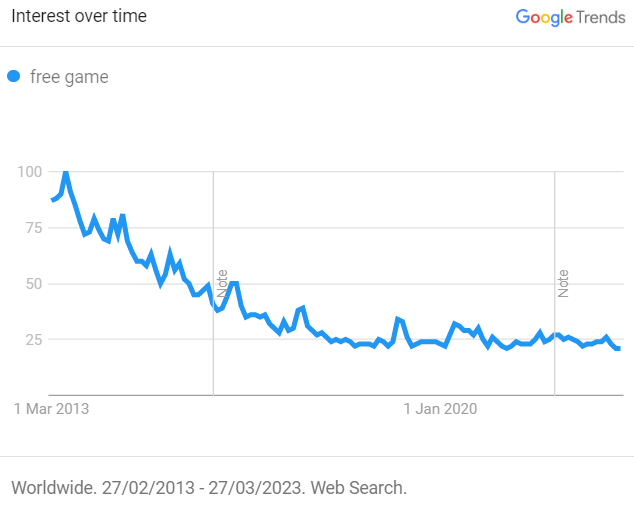}
\end{center}
    \caption{\small Google search queries for "free game" \cite{google_trends_game}.}
    \label{fig:game_google_trends}
\end{figure}

Despite negativity, there does remain a sustained interest in blockchain games, albeit in decline since the pandemic. Figure \ref{fig:bg_google_trends} illustrates this in the form of Google Trends search I performed, showing a spike in 2017 around the Cryptokitties hype \cite{cryptokitties_was_a_bubble}, and again in 2021-2022 around move-to-earn and Axie Infinity hype \cite{axie_infinity_was_a_bad_p2e_bubble}. For comparison, I list this against the overall trend for the "free game" search term, which has been in decline since 2013 (Figure \ref{fig:game_google_trends}).

\subsection{Environmental Impact}
It is worth addressing the environmental cost of blockchain-games too, given the infamously damaging effect cryptocurrencies have had on the environment. For instance, in 2022 it was estimated that Bitcoin consumes $\sim$100 TWh (360 PJ) annually, which equates to roughly as much electricity usage as Egypt \cite{btc-energy-consumption}. While Ethereum finally switched to a proof-of-stake model in 2022 \cite{eth_is_pos}, it was the blockchain-of-choice throughout the 2017-2022 prime of blockchain games, the environmental cost of which may be felt for a long time to come.

For example, according to Offsetra, the average ETH transaction in May 2021 created 18.05kg of CO2 equivalent emissions \cite{eth_pos_carbon_cost, eth-pow-carbon-cost, ubi_environment_analysis}. For the Sorare token, which saw over 673K transactions in Ubisoft's NFT-based BG \textbf{OneShot League} by May 2021 \cite{oneshotleague_2023, ubi_environment_analysis}, we can extract an estimate of the game's emissions:
\begin{align*}
    18.05\text{kg} \times 673,000\text{T} & = \textbf{12,147,650\text{kg}}\text{ CO2 Equivalent} \\
    & = 159 \text{ tanker trucks' worth of gasoline \cite{greenhouse_gas_measurements}.}
\end{align*}

By contrast, as of 2023/03/27, the Ethereum footprint is estimated at a much more tenable 0.02kg CO2e per transaction \cite{eth_pos_carbon_cost}: an $(18.05-0.02)/0.02=99.88\%$ reduction. While a move of Bitcoin from proof-of-work to proof-of-stake should also technically be possible, it is unclear whether this idea will ever take hold of enough maintainers for the fork to occur, with some seeing it as a risky violation of Nakamoto's original intentions for a stable blockchain \cite{bitcoin_to_pos}. This is not to mention the fact that miners are inherently incentivised by payment to keep the model proof-of-work.

\begin{figure}[h]
\begin{center}
\includegraphics[width=1\linewidth]{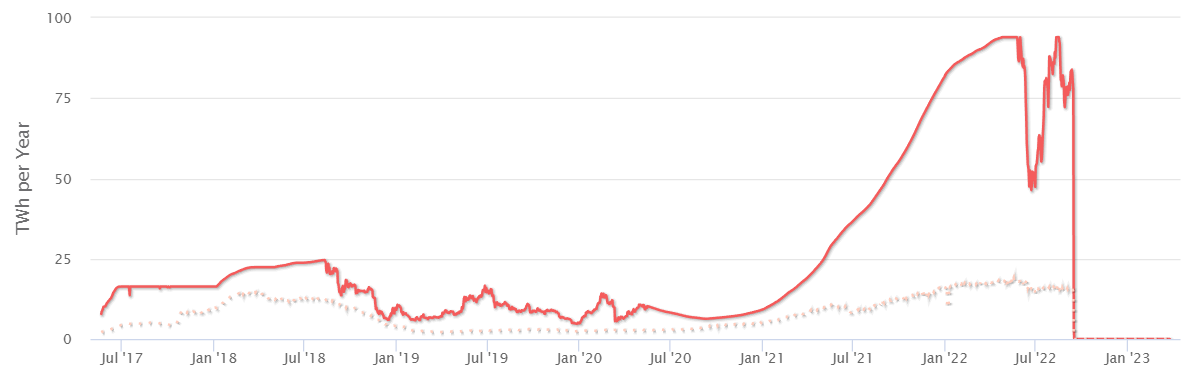}
\end{center}
    \caption{\small Ethereum's energy usage according to EthereumEnergyConsumption.com \cite{eth_energy_consumption_drop}; notice the complete drop at the introduction of proof-of-stake in 2022.}
    \label{fig:eth_energy_chart}
\end{figure}

The impact of Ethereum's switch to PoS is illustrated in Figure \ref{fig:eth_energy_chart}.

\subsection{Transaction Fees}
In general, blockchain games do not seem to be unduly limited by transaction fees. While high gas costs on the Ethereum blockchain have been problematic during times of congestion (such as in late 2021, when average transaction fees were as high as \$50 for a number of months \cite{eth_transaction_fees}), nowadays alternatives like the Ronin sidechain \cite{ronin} enable negligible transaction costs, and Etherum has at least maintained under \$5 since switching to PoS \cite{eth_transaction_fees}. Delegated Proof-of-Stake (DPoS)-based blockchain EOS attained popularity in 2018 for its zero gas fees, becoming particularly prevalent amongst gambling games \cite{min2019blockchain}.

\section{Discussion}

\subsection{Non-Blockchain Game Markets}
I believe that at a fundamental level, the only intrinsic gain to be had from making use of a \textit{blockchain} implementation rather than a \textit{centralised} one, for any particular system, is an enhanced level of \textbf{trust} \cite{nakamoto_2008} \cite{buterin2016ethereum}. While blockchain systems can help set up interoperability of assets across games, transparency of transactions, and scalability, these features could all be recreated by centralised means.

For example, Valve's 2007 first-person-shooter \textbf{Team Fortress 2} \cite{tf2} has operated an in-game economy since 2010, in which players receive and trade items with randomised stats. While inter-player trading for real-world currency initially took place exclusively on third-party grey markets, in 2012 Valve centralised the process with the introduction of the \textit{Steam Community Market}, to counteract fraud and provide a level of trust (well, and to take a share of the trade costs). PC Gamer described this large-scale, transparent marketplace as the "groundwork" for NFTs \cite{steam_birthed_nfts}.

\begin{figure}[h]
\begin{center}
\includegraphics[width=0.48\linewidth]{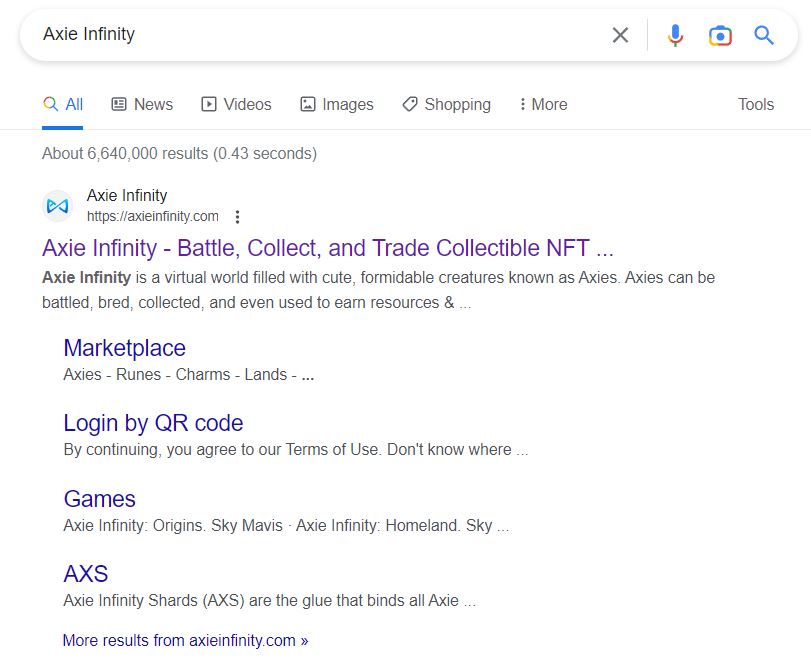}
\hfill
\includegraphics[width=0.48\linewidth]{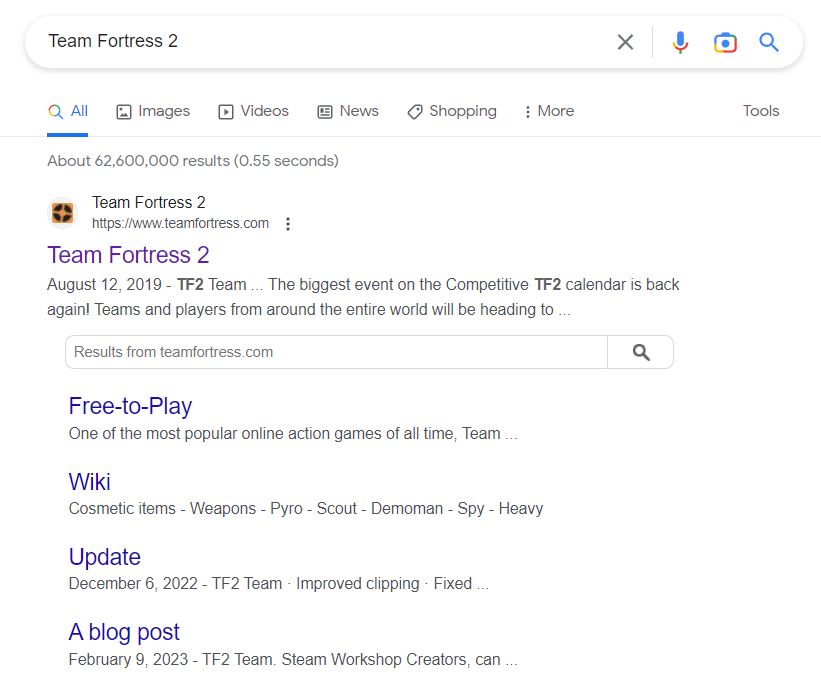}
\end{center}
    \caption{\small Top Google search result (2023/03/25) for Axie Infinity \cite{axie-infinity} and Team Fortress 2 \cite{tf2}. Notice Axie Infinity's use of "NFT" in the website title. Even more striking examples would be \textit{Cryptokitties} \cite{cryptokitties_2023} and \textit{Decentraland} \cite{decentraland_2023}, in which the blockchain integration is indicated in the names of the games themselves.}
    \label{fig:axie_vs_tf2}
\end{figure}

So, if we are to believe that blockchain integration is a merely a boon for smaller games to achieve the level of economic trust seen by larger, reputable companies such as Valve, it becomes clear why their branding is so different. Despite non-blockchain economies like Valve's TF2 and CS:GO being valued at tens or hundreds of millions of dollars \cite{tf2_50_million} \cite{csgo_1_billion}, their branding tends to downplay the presence of this in-game market, whereas I found that BGs will almost invariably put it front-and-centre. I illustrate this with an example in Figure \ref{fig:axie_vs_tf2}.

In review, it is clear why this marketing push exists: blockchain is a buzzword \cite{blockchain-is-a-buzzword} which excites buyers and empowers them with a sense of trust. This might explain why centralised in-game markets such as TF2's and CSGO's see top sales in order of thousands of dollars \cite{tf2_expensive_items}, while individual NFT sales have reached tens of millions \cite{nfts-are-expensive} despite NFT markets having such a shorter history.

\subsection{Conclusion}
In conclusion, after investigating the genre, I find several characteristics to be common amongst modern blockchain games:
\begin{itemize}
    \item Incorporation of multiple forms of blockchain integration, with a focus on \textbf{play-to-earn} and \textbf{marketplace} dynamics (Section \ref{sec:play-to-earn}, \ref{sec:ingame-markets}).
    \item An emphasis on blockchain as a means of investing in \textbf{digital assets} (Section \ref{sec:nontoken}).
    \item A \textbf{bubble-like} popularity that reacts to market trends, not just player preferences (Section \ref{sec:existing}).
\end{itemize}
While environmental concerns and transaction fee worries have some merit, the key limiter in the success of blockchain-integrated games appears to be public opinion. Crossing the barrier from the crypto-investment community into the wider gaming community is a challenge that I do not believe has yet been crossed, with recent attempts such as Ubisoft's Quartz meeting disastrous results \cite{ubi_quartz_bad, ubi_quartz_failed}.

While I remain optimistic for blockchain as a whole, and do believe that there is untapped potential in its overlap with gaming (in particular for the purpose of establishing long-lasting \textit{interoperability}, which is challenging in a centralised context due to gaming platforms being so ephemeral), I am sceptical of the blockchain-gaming landscape as it exists today. With so many games appealing more to investors than players, and using a monetization model closer to Ponzi schemes than traditional gaming setups, it is important to approach blockchain-gaming with caution.

\newpage
\appendix
\section{Quoted Reasons for Blockchain Integration}
\begin{itemize}
    \item "Blockchain's decentralised technology...changes the videogame industry by introducing concepts like uniqueness" \cite{ubisoft_quartz}.
    \item "[it's] about giving people options and helping them do more on their devices and in the cloud" \cite{microsoft_accepts_btc}.
    \item "I think that in the context of the games we create and the live services that we offer, collectible digital content is going to play a meaningful part in our future," \cite{ea_likes_nfts}.
    \item "We will...increase the possibilities of games, and reach out to users all over the world. We will continue to create new moving experiences." \cite{sega_likes_nfts}.
\end{itemize}

\bibliographystyle{plainjournalurl}
\bibliography{main}

\end{document}